\documentclass[twoside,twocolumn,9pt]{article}
\usepackage{extsizes}
\usepackage[super,sort&compress,comma]{natbib} 
\usepackage[version=3]{mhchem}
\usepackage[left=1.5cm, right=1.5cm, top=1.785cm, bottom=2.0cm]{geometry}
\usepackage{balance}
\usepackage{mathptmx}
\usepackage{sectsty}
\usepackage{graphicx} 
\usepackage{lastpage}
\usepackage[format=plain,justification=justified,singlelinecheck=false,font={stretch=1.125,small,sf},labelfont=bf,labelsep=space]{caption}
\usepackage{float}
\usepackage{fancyhdr}
\usepackage{fnpos}
\usepackage[english]{babel}
\addto{\captionsenglish}{%
  
}
\usepackage{array}
\usepackage{droidsans}
\usepackage{charter}
\usepackage[T1]{fontenc}
\usepackage[usenames,dvipsnames]{xcolor}
\usepackage{setspace}
\usepackage{hyperref}
\usepackage[compact]{titlesec}
\usepackage{hyperref}
\usepackage{xcolor}
\usepackage[switch]{lineno}

\usepackage{epstopdf}%This line makes .eps figures into .pdf - please comment out if not required.

\definecolor{cream}{RGB}{222,217,201}

\begin{document}

\pagestyle{fancy}
\thispagestyle{plain}
\fancypagestyle{plain}{
%%%HEADER%%%
\renewcommand{\headrulewidth}{0pt}
}
%%%END OF HEADER%%%

%%%PAGE SETUP - Please do not change any commands within this section%%%
\makeFNbottom
\makeatletter
\renewcommand\LARGE{\@setfontsize\LARGE{15pt}{17}}
\renewcommand\Large{\@setfontsize\Large{12pt}{14}}
\renewcommand\large{\@setfontsize\large{10pt}{12}}
\renewcommand\footnotesize{\@setfontsize\footnotesize{7pt}{10}}
\makeatother

\renewcommand{\thefootnote}{\fnsymbol{footnote}}
\renewcommand\footnoterule{\vspace*{1pt}% 
\color{cream}\hrule width 3.5in height 0.4pt \color{black}\vspace*{5pt}} 
\setcounter{secnumdepth}{5}

\makeatletter 
\renewcommand\@biblabel[1]{#1}            
\renewcommand\@makefntext[1]% 
{\noindent\makebox[0pt][r]{\@thefnmark\,}#1}
\makeatother 
\renewcommand{\figurename}{\small{Fig.}~}
\sectionfont{\sffamily\Large}
\subsectionfont{\normalsize}
\subsubsectionfont{\bf}
\setstretch{1.125} %In particular, please do not alter this line.
\setlength{\skip\footins}{0.8cm}
\setlength{\footnotesep}{0.25cm}
\setlength{\jot}{10pt}
\titlespacing*{\section}{0pt}{4pt}{4pt}
\titlespacing*{\subsection}{0pt}{15pt}{1pt}
%%%END OF PAGE SETUP%%%

%%%FOOTER%%%
\fancyfoot{}
\fancyfoot[LO,RE]{\vspace{-7.1pt}\includegraphics[height=9pt]{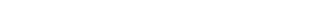}}
\fancyfoot[CO]{\vspace{-7.1pt}\hspace{11.9cm}\includegraphics{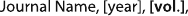}}
\fancyfoot[CE]{\vspace{-7.2pt}\hspace{-13.2cm}\includegraphics{head_foot/RF}}
\fancyfoot[RO]{\footnotesize{\sffamily{1--\pageref{LastPage} ~\textbar  \hspace{2pt}\thepage}}}
\fancyfoot[LE]{\footnotesize{\sffamily{\thepage~\textbar\hspace{4.65cm} 1--\pageref{LastPage}}}}
\fancyhead{}
\renewcommand{\headrulewidth}{0pt} 
\renewcommand{\footrulewidth}{0pt}
\setlength{\arrayrulewidth}{1pt}
\setlength{\columnsep}{6.5mm}
\setlength\bibsep{1pt}
%%%END OF FOOTER%%%

%%%FIGURE SETUP - please do not change any commands within this section%%%
\makeatletter 
\newlength{\figrulesep} 
\setlength{\figrulesep}{0.5\textfloatsep} 

\newcommand{\topfigrule}{\vspace*{-1pt}% 
\noindent{\color{cream}\rule[-\figrulesep]{\columnwidth}{1.5pt}} }

\newcommand{\botfigrule}{\vspace*{-2pt}% 
\noindent{\color{cream}\rule[\figrulesep]{\columnwidth}{1.5pt}} }

\newcommand{\dblfigrule}{\vspace*{-1pt}% 
\noindent{\color{cream}\rule[-\figrulesep]{\textwidth}{1.5pt}} }

\makeatother
%%%END OF FIGURE SETUP%%%

%%%% For Review %%%%%%%

%%%TITLE, AUTHORS AND ABSTRACT%%%
\twocolumn[
  \begin{@twocolumnfalse}
{\includegraphics[height=30pt]{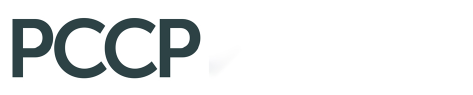}\hfill\raisebox{0pt}[0pt][0pt]{\includegraphics[height=55pt]{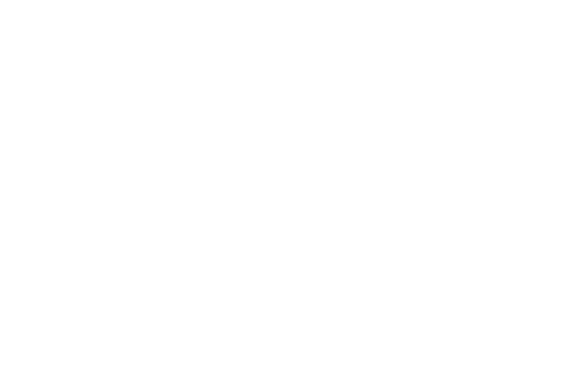}}\\[1ex]
\includegraphics[width=18.5cm]{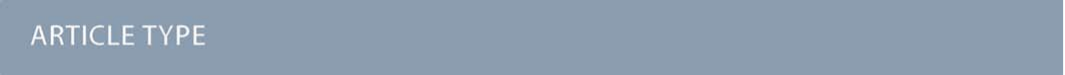}}\par
\vspace{1em}
\sffamily
\begin{tabular}{m{4.5cm} p{13.5cm} }

\includegraphics{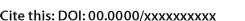} & \noindent\LARGE{\textbf{Quantifying Errors in Chemical Kinetics$^\dag$}} \\
\vspace{0.3cm} & \vspace{0.3cm} \\

 & \noindent\large{Skyla B.\ White,$^{\ast}$\textit{$^{a,}$}\textit{$^{b}$} Paul B.\ Rimmer,\textit{$^{a}$} Miles Cramner\textit{$^{c,d}$} and Alex T.\ Archibald \textit{$^{e,f}$}}\\

\includegraphics{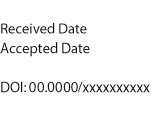} & \noindent\normalsize{The physical structure of kinetic measurements determines the correct statistical framework for quantifying their uncertainties, yet this framework has never been explicitly derived or systematically applied in the chemical kinetics literature. Here, we derive a forward model for kinetic observations from a minimal set of stated assumptions about the measurement process, and identify the correct statistical treatment for each kinetic parameter as an explicit consequence of that model. We then audit the literature pertaining to a well-studied reaction, examining the kinetics of permanganate oxidation, to establish the scale of deviation from the correct framework. We find that of 330 measurements, 35.8\% carry no uncertainty quantification of any kind and none are accompanied by a full propagation of all uncertainties. To demonstrate the practical consequences of departing from the correct framework, we reanalyse a representative kinetic dataset, first checking whether the data are consistent with the forward model and then applying the correct procedures. We find that the impacts are modest for the dataset examined, as expected for quantities determined in a low-uncertainty regime, but that uncertainties in activation energy are systematically underestimated when observed rather than assumed measurement precision is used as the basis for regression weights. We highlight that these consequences become substantially more significant at the limits of current experimental methodologies, where measurement uncertainty is large and the correct distributional treatment can no longer be approximated by standard approaches. Based on this, we propose a set of minimum recommendations for error reporting in chemical kinetics that follow directly from our forward model.} \\

\end{tabular}

 \end{@twocolumnfalse} \vspace{0.6cm}

  ]
%%%END OF TITLE, AUTHORS AND ABSTRACT%%%

%%%FONT SETUP - please do not change any commands within this section
\renewcommand*\rmdefault{bch}\normalfont\upshape
\rmfamily
\section*{}
\vspace{-1cm}

%%%FOOTNOTES%%%

\footnotetext{\textit{$^{a}$~Cavendish Laboratory, University of Cambridge, JJ Thomson Avenue, Cambridge CB3 0US, United Kingdom}}
\footnotetext{\textit{$^{b}$~Current Address: Center for Space and Habitability (CSH)}, University of Bern, Gesellschaftsstrasse 6, 3012 Bern, Switzerland}
\footnotetext{\textit{$^{c}$~Department of Applied Mathematics and Theoretical Physics, University of Cambridge, Wilberforce Road, Cambridge CB3 0WA, United Kingdom}}
\footnotetext{\textit{$^{d}$~Institute of Astronomy, University of Cambridge, Madingley Road, Cambridge CB3 0HA, United Kingdom}}
\footnotetext{\textit{$^{e}$~Yusuf Hamied Department of Chemistry, University of Cambridge, Cambridge, CB2 1EW, United Kingdom}}
\footnotetext{\textit{$^{f}$~National Centre for Atmospheric Science, University of Cambridge, Cambridge, United Kingdom}}

%Please use \dag to cite the ESI in the main text of the article.
%If you article does not have ESI please remove the the \dag symbol from the title and the footnotetext below.
%\footnotetext{\dag~Supplementary Information available: [details of any supplementary information available should be included here]. See DOI: 10.1039/cXCP00000x/}
%additional addresses can be cited as above using the lower-case letters, c, d, e... If all authors are from the same address, no letter is required

\footnotetext{\textit{$^{*}$~Corresponding Author: Skyla B. White \\ Email: skyla.white@unibe.ch}}

%%%END OF FOOTNOTES%%%

%%%MAIN TEXT%%%%

\section*{Introduction}

Chemical kinetics has long played a critical role in understanding and predicting the behaviour of complex dynamic systems. In pharmaceutical science, kinetic models inform predictions of optimal antibiotic treatment strategies, directly guiding clinical decisions \cite{Abel2017}. In climate science, they support the modelling of biogeochemical cycles, shaping policy discussions on how best to address climate change \cite{Wang2007}. Similarly, in origin of life research, kinetics guides experiments on the formation of life's building blocks, identifying environments that may favour particular chemical pathways \cite{White2024b}. More recently, advances in computational methods and infrastructure have made it increasingly feasible to couple kinetic frameworks with large-scale computational models, dramatically expanding the complexity and ambition of the systems kinetics can underpin \cite{Grinberg2026}. Rather than serving solely as a descriptive framework, kinetics now drives decision making across diverse scientific domains at a scale that was previously out of reach.

As kinetics is embedded in models of growing scale and consequence, the accuracy of the underlying data becomes critically important. Uncertainty has always mattered in kinetics, but the way in which kinetic data is now used has changed the stakes considerably. Large-scale predictive models increasingly draw on kinetic parameters that were often reported as a byproduct of studies with other primary goals, and as such, were never intended to be extracted and reused in this way. The uncertainty standards appropriate for the original context are not necessarily sufficient for this new one. Despite this, rigorous statistical treatment of kinetic results has not yet become standard practice \cite{Vsimon2022}. This is perhaps unsurprising, reflecting a combination of factors. First, rapid publication cycles make careful uncertainty reporting difficult to prioritise \cite{Ciriminna2024}; second, the guidelines that do exist carry no enforcement mechanism within the primary literature \cite{Bazyleva2021, Bruno2021, Shaw-2023, eurachem1995_quantifying_uncertainty, Verma2025}; third, reporting conventions in kinetics have not changed much over the decades with authors naturally following the established norms of their field \cite{Cvetanovic1979}; and finally, it is worth acknowledging that more rigorous uncertainty reporting will not produce better values, instead it will simply reveal how uncertain the existing ones are, and in many cases uncertainties will appear larger than previously reported. In the absence of any universal standards this ``inflation'' of error is a legitimate concern for many researchers. The consequence of this lack of rigorous statistical treatment is that studies reporting kinetic data often omit uncertainty quantification or apply inconsistent methods for estimating uncertainties for equivalent parameters. As a result, reported values are not directly comparable across studies. Moreover, this limits their utility in models, as uncertainty bounds may be absent or inconsistently defined, preventing predictions from capturing the full range of possible outcomes. Importantly, knowing the true uncertainty of a measurement is a prerequisite for reducing it, whether through improved experimental design, better instrumentation, or more appropriate computational methods.

The concept of error reporting has long been recognised across scientific disciplines \cite{Taylor2022, JCGM100:2008}, yet in chemical kinetics there remains little consensus on how uncertainties should be quantified or communicated. Kinetic parameters may be obtained through a variety of experimental and computational approaches \cite{Fernandez2006}, each introducing distinct sources of uncertainty and requiring different strategies for error propagation \cite{Goldsmith2013, Franco2015, Wang2015}. Although a wide range of methods for uncertainty quantification have been developed, and significant progress has been made in computational studies to assess and mitigate the resulting uncertainties \cite{Wakelam2010, Loison2015, Hebrard2015}, these approaches are typically applied in a fragmented manner and do not explicitly account for the underlying physical structure of kinetic measurements, that is, the chain of operations through which experimental observations are converted into reported kinetic parameters. A good example of this is the now widely adopted use of Monte Carlo and bootstrapping methods, which are often seen as a robust way to obtain uncertainties without deriving them analytically. However, Monte Carlo propagation requires the distribution of each input to be specified, and bootstrapping, although distribution-free, can only recover the underlying distribution from samples large enough to contain that information \cite{Hesterberg2015}, which kinetic datasets rarely provide. In addition, the outcome of either method must still be summarised and reported, and an appropriate method does not ensure an appropriate representation of its result. Consequently, current reporting practices lack a common, physically grounded statistical framework.

The physical structure of kinetic measurements is not arbitrary, and neither therefore is the correct statistical framework for quantifying their uncertainties. Kinetic parameters are derived quantities whose distributional properties follow from how they are calculated and what sources of uncertainty act on them. Some parameters are log-normally distributed, others are normally distributed, and the correct choice of estimator, the form of reported uncertainty bounds, and the appropriate regression procedure all differ accordingly \cite{Andersson2021, Motulsky2025}. This framework can be derived by specifying a forward model for how experimental observations are generated, and the statistical procedures that follow from it are consequences of that model. Despite this, the literature does not reflect this framework. In gas-phase atmospheric chemistry, evaluation panels such as the NASA Jet Propulsion Laboratory (JPL) \cite{Burkholder2020} and the International Union of Pure and Applied Chemistry (IUPAC) \cite{Mellouki2021} have implicitly recognised this structure by reporting uncertainties as multiplicative factors in log-space, though these panels are mutually inconsistent in the probabilistic meaning of their factors, and both explicitly acknowledge that their uncertainties represent expert judgement rather than formal statistical analysis \cite{Atkinson2006, Burkholder2020}. Furthermore, this practice is not consistently adopted in the underlying primary literature \cite{Mcgillen2020}, nor in domains such as aqueous-phase kinetics, where no equivalent evaluated database exists. Across the broader kinetics literature, symmetric additive intervals remain the norm, implicitly assuming normality regardless of whether that assumption is justified, and regression analyses typically rely on ordinary least squares (OLS) without verifying that measurement uncertainties are consistent with observed scatter. These deviations from correct practice produce uncertainty bounds that do not honestly represent the range of plausible values, and when propagated into predictive models produce overconfident outputs.

Historically, the impact of these deviations has been partially mitigated by expert interpretation. Researchers can assess methodological context, recognise anomalous values, and apply informal judgement when evaluating published data. However, this safeguard diminishes in the context of predictive modelling and large-scale data extraction, particularly with the rise of large language models \cite{Johnson2022, Komp2022, Yano2022}. These systems lack the domain intuition required to evaluate experimental limitations or methodological caveats. When kinetic parameters are reported without uncertainty bounds or with inappropriate assumptions, large language models may treat them as exact quantities. The resulting propagation of mischaracterised uncertainty through predictive frameworks risks producing misleading results making rigorous reporting of kinetic uncertainties a necessity.

To date, there exists no single regulatory authority for chemical kinetics error reporting. Instead, reporting practices are shaped by various bodies. For example, IUPAC provide guidelines for how to report chemical data \cite{Bazyleva2021}, detailing the importance of completeness, the inclusion of metadata, and abiding by FAIR (findable, accessible, interoperable, and reusable) data sharing \cite{Bruno2021, Shaw-2023}. In addition, there exist guides such as those provided by the Joint Committee for Guides in Metrology (JCGM) and the National Institute of Standards and Technology (NIST) \cite{eurachem1995_quantifying_uncertainty, Possolo2015} that define the standard principles for reporting measurement errors and uncertainty. There are also quality frameworks such as Good Laboratory Practice (GLP) which play a role in the quantification of measurement uncertainty by providing a structured framework that ensures data is reliable \cite{Verma2025}. The most developed kinetics-specific efforts come from the NASA/JPL Panel for Data Evaluation \cite{Burkholder2020} and the IUPAC Task Group on Atmospheric Chemical Kinetic Data Evaluation \cite{Atkinson2006} as detailed previously. Finally, there exist numerous papers discussing this topic at length \cite{Cvetanovic1979, Pardue1989, Hartland2020}. Despite these efforts, many papers containing kinetic data include no discussion of errors, and even those that do rarely adopt a statistically consistent framework.

The extent to which current practice deviates from the statistically correct framework has not been systematically characterised, nor has that framework itself been explicitly derived for the full range of kinetic parameters encountered in practice. This work addresses both gaps. We first derive a forward model for kinetic observations from the physical structure of the measurement process, and identify the correct statistical procedures for each kinetic parameter as consequences of that model. We then audit current error reporting practices using a large, independently compiled dataset of rate constants for permanganate oxidation of organic compounds, establishing the scale of the deviation from correct practice across 57 years of literature. We then apply the correct procedures derived from the forward model to a representative kinetic dataset, demonstrating the practical consequences of adopting the appropriate statistical framework. From these findings we propose a set of minimum recommendations for error reporting in chemical kinetics that follow directly from the forward model.

\section*{Statistical Framework for Kinetic Parameters}

The correct statistical treatment of kinetic parameters follows from a generative model for how experimental observations are produced. In this section we derive this model from the physical structure of kinetic measurements and identify the correct statistical procedures for a variety of kinetic parameters as explicit consequences of it.

\subsection*{The Physical Basis for the Forward Model}

The central limit theorem states that the sum of many independent random variables converges to a normal distribution regardless of the distributions of the individual variables. An analogous result holds for products: since taking the logarithm converts a product into a sum, multiplying many independent uncertain quantities together produces a log-normal distribution. The key question for kinetic measurements is therefore whether uncertainty enters additively or multiplicatively, and the physical structure of kinetic experiments suggests the latter. Concentrations are prepared through weighing and dilution, both multiplicative operations, and instrumental precision typically scales with the quantity being measured rather than remaining constant in absolute terms. Kinetic parameters are then derived from these quantities through further multiplicative relationships. Where multiple sources of multiplicative uncertainty act together, the central limit theorem therefore provides theoretical grounding for treating many kinetic parameters as log-normally distributed. We emphasise that this is a well-motivated assumption grounded in the physics of the measurement process rather than a proven fact, and all statistical procedures derived below are conditional on it holding.

\subsection*{The Forward Model}

The rate constant is the fundamental quantity in kinetics from which all other parameters are either derived or to which they contribute. We therefore express the forward model in terms of $\ln(k)$, from which the distributional properties of other kinetic parameters follow as consequences. 

For each compound $j$, we assume there exists a true rate constant $k_j$ whose logarithm, $\ln(k_j)$, is a fixed physical quantity. Each experimental measurement of $k_j$ is treated as a noisy observation of this true value such that:
\begin{equation}
    \ln(k_{\text{obs},i,j}) \sim \mathcal{N}(\ln(k_j), \sigma^2_{\ln,j})
\end{equation}
where $i$ indexes independent replicates and $\sigma^2_{\ln,j}$ characterises the measurement noise for compound $j$. This is equivalent to assuming $k_{\text{obs}}$ is log-normally distributed around the true $k_j$. For measurements that span temperature $k_j(T)$ can be described by the Arrhenius equation:
\begin{equation}
    \ln(k_{j}(T)) = \ln(A_j) - \frac{E_{\text{a},j}}{RT}
\end{equation}
where $\ln(A_j)$ and $E_{\text{a},j}$ are per compound latent parameters and temperature $T_{t,j}$ is treated as a known observed quantity. $R$ is the universal gas constant. In this case the model extends to:
\begin{equation}
    \ln(k_{\text{obs},i,t,j}) \sim \mathcal{N}(\ln(A_j) - \frac{E_{\text{a},j}}{RT_{t,j}}, \sigma^2_{\ln,j})
\end{equation}

\begin{figure}[ht]
    \centering
    \includegraphics[width=1.0\linewidth]{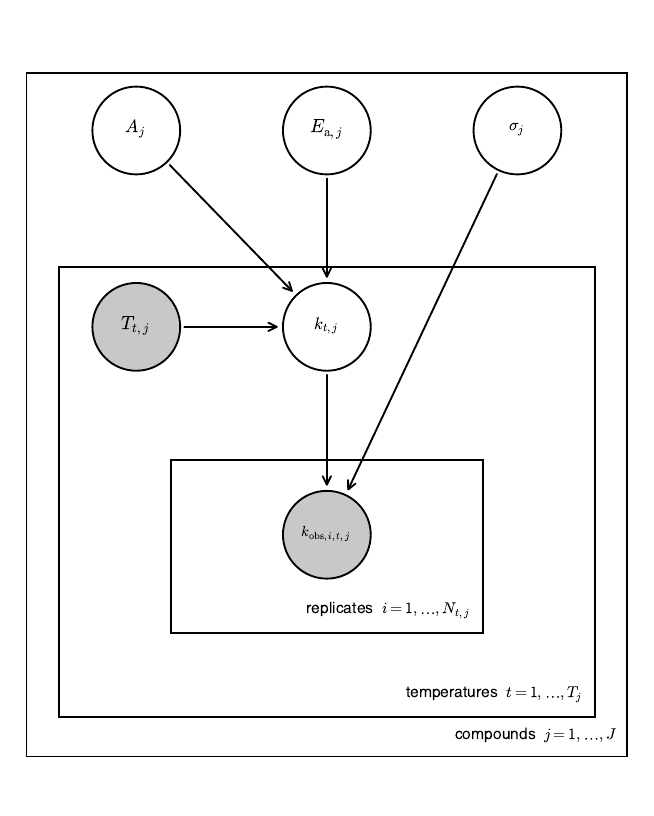}
    \caption{\textbf{Probabilistic Graphical Model.} The schematic describes the generative model for kinetic observations. White nodes are quantities the model estimates. Grey nodes are observed quantities: $T_{t,j}$ is a known experimental covariate, while $k_{\text{obs}, i,t,j}$ is the measured outcome the model is fitted to. Arrows indicate conditional dependence between variables i.e. here $k_{t,j}$ depends on $A_j, E_{\text{a},j}$ and $T_{t,j}$; and $k_{\text{obs}, i,t,j}$ depends on $\sigma_{j}$ and $k_{t,j}$. Plates denote repeated structure: all variables within the compounds plate exist once per compound $j$; all variables within the temperatures plate exist once per temperature $t$; all variables within the replicates plate exist once per replicate $i$. This model applies in full where temperature-dependent data are available. Where rate constants are determined at a single temperature only, the Arrhenius layer collapses and the model reduces to a single true $k_j$ per compound.}
    \label{pgm}
\end{figure}

Figure \ref{pgm} shows a probabilistic graphical model describing the relationships between variables in the above forward model. 

The join probability equation based on Figure \ref{pgm} is as follows:

\begin{multline}
    P = \prod_j \Bigg[ p(A_j) \cdot p(E_{\text{a},j}) \cdot p(\sigma_{j}) \\
    \cdot \prod_t \Bigg[ p(k_{t,j} \mid A_j, E_{\text{a},j}, T_{t,j}) 
    \cdot \prod_i p(k_{\text{obs},i,t,j} \mid k_{t,j}, \sigma_{j}) \Bigg] \Bigg]
\end{multline}

We note that the forward model as derived assumes that the temperature dependence of the rate constant is adequately described by the Arrhenius equation. This is itself a model rather than a physical law, and there are well-documented cases where it breaks down \cite{Carvalho2019}. In such cases, fitting a linear model to $\ln(k)$ versus $T^{-1}$ is physically inappropriate regardless of the statistical procedure used, and the uncertainty derived from such a regression would not meaningfully characterise the uncertainty in the underlying kinetic parameters. The statistical framework presented here is not itself dependent on the Arrhenius parameterisation and remains valid for any kinetic model in which uncertainty enters multiplicatively. However, for non-Arrhenius systems, the specific form of the regression model and the parameters to which uncertainties are assigned would need to be revised to reflect the appropriate physical description of the temperature dependence.

\subsection*{Consequences of the Forward Model}

Given this forward model, the correct statistical treatment of each kinetic parameter follows as a direct consequence. We derive these here explicitly.

Under the forward model, rates and rate constants are log-normally distributed, as they are either products of measured quantities or derived from them through multiplicative relationships. The maximum likelihood estimator of the true rate constant is therefore the geometric mean of replicate measurements:
\begin{equation}
    \bar{k}_\text{geo} = \exp \left( \frac{1}{n}\sum_{i=1}^{n}\ln(k_i) \right)
\end{equation}
where $n$ is the number of replicate measurements and $i$ indexes each individual value.

Since rates are products of concentrations and rate constants, and uncertainty enters through the same multiplicative mechanisms, the same forward model applies and the maximum likelihood estimator of the true rate is again the geometric mean, with an identical form to that given above for $k$, but expressed in terms of $r$.

The natural summary of spread for both rate constants and rates is then the standard deviation of the log-transformed measurements ($\sigma_{\ln}$), and uncertainties should be expressed as asymmetric bounds. Here, the standard deviation in logarithmic space  is calculated as:
\begin{equation}
    \sigma_{\ln} = \sqrt{\frac{1}{n-1}\sum_{i=1}^{n}\left(\ln(k_{,i}) - \ln(\bar{k}_\text{geo})\right)^2}
\end{equation}
An analogous expression applies for rates, with $r$ in place of $k$.

For the rate constant bounds are then expressed as:
\begin{equation}
    \text{upper bound} = \bar{k}_{\text{geo}} \cdot \exp(\sigma_{\ln})
\end{equation}
\begin{equation}
    \text{lower bound} = \bar{k}_\text{geo} \cdot \exp(-\sigma_{\ln})
\end{equation}
with equivalent expressions for rates obtained by replacing $k$ with $r$.

A related quantity is the reaction order, defined as the slope of a linear regression of $\ln(\text{rate})$ versus $\ln(\text{concentration})$. Under the forward model $\ln(\text{rate})$ is normally distributed, being the logarithm of a log-normal quantity, as it is a linear combination of normal random variables. The slope of a linear regression with normally distributed errors is itself normally distributed. Reaction orders are therefore normally distributed and symmetric error bounds are the correct representation. The following treatment assumes that the rate law follows a simple power law for which reaction order is a well-defined constant.

In addition, the forward model also determines how this regression should be performed. The forward model specifies that log-transformed kinetic quantities are normally distributed with variance $\sigma_{\ln}^2$. For any regression in which the dependent variable is a log-transformed kinetic quantity and in which $\sigma_{\ln}$ is known, the maximum likelihood estimator of the regression parameters is therefore weighted least squares (WLS) with weights $\omega = 1/\sigma_{\ln}^2$. Ordinary least squares is a special case that is only valid when $\sigma_{\ln}$ is constant across all data points, and should not be assumed without verification. This applies to both the determination of reaction orders and activation energies. 

The uncertainty in the parameter derived from this regression analysis and propagated from the regression weights reflects assumed measurement precision. Where $\sigma_{\ln}$ is estimated from replicates, the regression weights capture observed measurement variability and the resulting uncertainty is grounded in the data. However, even where weights are correctly specified, the observed scatter of data points about the regression line may exceed the scatter predicted by those weights, indicating additional sources of variability not captured by the forward model. The reduced chi-squared statistic provides a diagnostic check on this consistency. Where it exceeds unity the uncertainty should be scaled by $\sqrt{\chi^2_{red}}$ as a conservative correction. We note that this scaling is not a direct consequence of the forward model but a pragmatic safeguard against unmodelled variability.

The activation energy itself is also derived as the slope of linear regression of $\ln(k)$ versus $T^{-1}$. Under the forward model $\ln(k)$ is normally distributed  and therefore $E_\text{a}$, being a regression slope in log-space, is normally distributed and symmetric error bounds are correct. However, as with reaction orders, the maximum likelihood estimator of $E_\text{a}$ is weighted least squares regression as described above. 

Finally, the pre-exponential factor $A$ can be derived as the intercept of the same Arrhenius regression in log-space. Thus, $\ln(A)$ is normally distributed under the forward model. $A$ itself is therefore log-normally distributed, and asymmetric error bounds are correct, even though $E_\text{a}$ derived from the same regression carries symmetric bounds.

A summary of these parameters and their related distributions can be seen in Table \ref{distributions}.

\begin{table*}[ht]
\centering
\caption{\textbf{Statistical Treatment of Kinetic Parameters Under the Forward Model.}}
\label{distributions}
\small
\renewcommand{\arraystretch}{1.6}
\begin{tabular}{p{3.5cm} p{2.5cm} p{5cm} p{3.5cm}}
\hline
\textbf{Parameter} & \textbf{Distribution} & \textbf{Central Estimate} & \textbf{Uncertainty Bounds} \\
\hline
Rate                   & Log-normal & Geometric mean                  & Asymmetric \\
Rate constant          & Log-normal & Geometric mean                  & Asymmetric \\
Reaction order         & Normal     & WLS regression slope            & Symmetric  \\
Activation energy      & Normal     & WLS regression slope            & Symmetric  \\
Pre-exponential factor & Log-normal & Back-transform of WLS intercept & Asymmetric \\
\hline
\end{tabular}
\end{table*}
The mixed picture presented in Table \ref{distributions}, where different kinetic parameters require different statistical treatments even when derived from the same dataset, is a direct consequence of the forward model. It illustrates why the model must be specified explicitly before any analysis is performed, and why blanket application of symmetric normally distributed error bounds is incorrect for some parameters even when it happens to be correct for others.

\section*{Methods}

This study was carried out in two phases. First, error reporting practices in the chemical kinetics literature were assessed through a systematic audit of published rate constants. Second, the correct statistical procedures derived from the forward model were applied to a published kinetic dataset, allowing direct comparison between results obtained under the correct framework and those reported under standard practice.

\subsection*{Literature Audit}

To establish the scale of deviation from the statistical framework derived above, an audit of the chemical kinetics literature was undertaken, in which publications reporting rate constants for permanganate oxidation of organic compounds were systematically reviewed. These papers were previously compiled by Laszakovits et al.\ \cite{Laszakovits2022}. This dataset was selected as it represents a comprehensive, independently assembled collection spanning 57 years of research (1965--2022), drawing from 90 distinct publications and encompassing 330 rate constant determinations for 216 unique organic compounds. For each entry, the original publication was inspected to identify how experimental uncertainty was quantified and communicated. Where Laszakovits et al.\ averaged values from several studies, or derived a second-order rate constant ($k_2$) from a reported pseudo-first-order rate constant ($k_{\text{obs}}$), the resulting uncertainty was not attributed to the original publication; instead, the uncertainty reported in the original publication on either $k_2$ or $k_{\text{obs}}$ was assessed.

Each measurement was assigned to one of six categories: (i) no error reported, where neither $k_2$ nor $k_{\text{obs}}$ was given a quantified uncertainty, including cases where only a blanket statement was made (e.g.\ ``standard deviations were usually within 5\%''); (ii) inaccessible or unverifiable, where the original publication could not be obtained or the value reported by Laszakovits et al.\ could not be located within it or its supporting information; (iii) error given, origin not described, where a numerical uncertainty was stated but nothing was said about its type, the number of repeats, or how it was obtained; (iv) error partially described, where some but not all of these three elements were stated. For example, Chen et al.\ \cite{Chen2005} report $k_{\text{obs}}$ values with uncertainties for $n = 10$ and derive $k_2$ from them, but do not state whether the uncertainty is a standard deviation, standard error, or confidence interval; (v) error fully described, where the type of uncertainty, the number of repeats, and the method were all stated in the text, a table, or a caption. For example, Hossain \& McLaughlan \cite{Hossain2012} report $k_{\text{obs}}$ for chlorophenol oxidation as the mean $\pm$ 95\% confidence interval of triplicate experiments, and calculate the second-order rate constant as $k_{\text{app}} = k_{\text{obs}}/[\mathrm{KMnO_4}]_0$ from the mean $k_{\text{obs}}$; notably, no uncertainty is given for $k_{\text{app}}$ itself; and (vi) full uncertainty treatment, where in addition to (v) all relevant sources of uncertainty (e.g.\ reagent concentrations, regression, and replication) were propagated into the reported $k_2$. Information that had to be inferred, for example by counting data points in a figure, was not considered stated.

Although some subjectivity was unavoidably introduced by the varied and often incomplete manner in which uncertainties were described in the original publications, the full basis for each assignment, including relevant quotations and any discrepancies with the values reported by Laszakovits et al.\, is provided in the Supporting Information so that the categorisation can be independently assessed.

\subsection*{Statistical Reanalysis}

The statistical procedures derived from the forward model were applied to a dataset reported by Huang et al.\ \cite{Huang2001} as a representative example of standard practice in the chemical kinetics literature. This dataset was selected because it is typical of experimental kinetic studies in its scope and methodology, and because it provides sufficient replicate measurements to permit empirical consistency checks against the forward model, albeit with limited statistical power given the small number of replicates per compound. These checks are therefore reported as indicative rather than definitive and are used to establish that nothing in the data contradicts the forward model, rather than to prove that the model is correct.  The results obtained under the correct framework were then compared with those originally reported by Huang et al., allowing the practical consequences of departing from the correct statistical framework to be assessed directly.

\subsubsection*{The Dataset}

 Data was taken from a study by Huang et al.\ \cite{Huang2001}. This study was focused on determining kinetic parameters for permanganate oxidation of chlorinated ethylenes. In this work they identified rate constants across the temperature range 10--25$^\circ$C for the reaction of permanganate with tetrachloroethylene (PCE), trichloroethylene (TCE), cis-1,2-dichloroethylene (cis-DCE), trans-1,2-dichloroethylene (trans-DCE), and 1,1-dichloroethylene (1,1-DCE). To determine rate constants they performed time-course experiments with permanganate in excess to ensure pseudo-first order kinetics with respect to the chlorinated ethylene. From these experiments they were able to determine $k_{\text{obs}}$. Second order rate constants were then determined from the relationship $k_{\text{obs}} = k_2[\rm KMnO_4]_0$. $k_2$ values were obtained for a variety of compounds, as listed above, with values at 20$^\circ$C being measured across a variety of permanganate concentrations. The resulting dataset comprises 27 rate constant measurements. They then further reported activation energies from linear regression analysis of plots of $\ln(\text{rate})$ against $T^{-1}$ for each of the above compounds following experiments conducted at 10, 15, 20, and 25 $^\circ$C.

 \subsubsection*{Distributional Analysis}

 The consistency of the Huang et al.\ dataset with the forward model was assessed empirically using three diagnostic metrics computed from replicate measurements of $k_{\text{obs}}$ and thus repeated determinations of $k_2$ at 20$^\circ$C. These metrics were the absolute standard deviation of $k_2$ ($\sigma_{k_2}$), the coefficient of variation (CV), and the standard deviation in log-space ($\sigma_{\ln(k_2)}$). Under the forward model, uncertainty enters multiplicatively and therefore CV and $\sigma_{\ln(k_2)}$ should remain approximately constant across compounds. If uncertainty entered additively instead, the absolute standard deviation would be expected to remain constant. The pattern of variation across these three metrics therefore provided evidence about whether the error mechanism was consistent with the forward model.

 CV was calculated as:
 \begin{equation}
    CV = \frac{\sigma_{k_2}}{k_2}
\end{equation}
where $k_2$ is the rate constant and $\sigma_{k_2}$ is the absolute standard deviation across replicate measurements at 20$^\circ$C. The relationship between CV and $\sigma_{\ln(k_2)}$ is given by:
\begin{equation}
    CV = \sqrt{\exp(\sigma_{\ln(k_2)}^2) - 1}
\end{equation}
which simplifies to:
\begin{equation}
    CV \approx  \sigma_{\ln(k_2)}
\end{equation}
when $\sigma_{\ln(k_2)}$ is small. Constant $\sigma_{\ln(k_2)}$ therefore would imply approximately constant CV, and both provide equivalent tests of consistency with the forward model. For a derivation of these parameters and relationships, the reader is referred to the Supporting Information.

Formal normality tests such as Shapiro-Wilk and graphical assessments such as Q-Q plots were not applied to individual compounds due to insufficient statistical power at $n = 5-6$ replicates per compound. The across-compound comparison of these three metrics was therefore used as an indirect consistency check. This approach is valid provided the measurement procedure was consistent across compounds, which was satisfied here given that all five compounds were measured under similar experimental conditions as detailed by Huang et al\cite{Huang2001}.

Following this consistency check, the correct estimators derived from the forward model were applied to the Huang et al.\ data. For each compound the geometric mean and $\sigma_{\ln(k_2)}$ were calculated at 20$^\circ$C. Asymmetric uncertainty bounds were then calculated as defined by the forward model. These were compared with the arithmetic means and symmetric confidence intervals originally reported by Huang et al.\ to illustrate the practical consequences of applying the correct framework versus standard practice. 

\subsubsection*{Regression Analysis}

Following the forward model, activation energies were determined from the Huang et al.\ dataset using weighted least squares regression of $\ln(k_2)$ versus $T^{-1}$, with weights $\omega_i = 1/\sigma_{\ln(k_2)}^2$ derived from replicate measurements at each temperature. These uncertainties were assumed constant across temperatures within each compound. This is a contrast to the ordinary least squares regression used by Huang et al., which is only the correct estimator under the forward model when $\sigma_{\ln(k_2)}$ is constant across all data points and equal for all compounds.

Six weighted moments were computed to support the regression. The total sum of weights was calculated as:

\begin{equation}
    S = \sum_{i=1}^{N} \omega_i
\end{equation}

where $N$ is the number of data points i.e.\ 4, one for each temperature $T$; and $i$ indexes the four measurement temperatures. The weighted sum of the inverse temperature values was calculated as:

\begin{equation}
    S_T = \sum_{i=1}^{N} \omega_i \cdot x_i
\end{equation}

where $x_i = 1/T_i$. The weighted sum of squared inverse temperature values was calculated as:

\begin{equation}
    S_{TT} = \sum_{i=1}^{N} \omega_i \cdot x_i^2
\end{equation}

The weighted sum of the log-transformed rate constants was calculated as:

\begin{equation}
    S_R = \sum_{i=1}^{N} \omega_i \cdot y_i
\end{equation}

where $y_i = \ln(k_{2,i})$. The weighted sum of squared log-transformed rate constants was calculated as:

\begin{equation}
    S_{RR} = \sum_{i=1}^{N} \omega_i \cdot y_i^2
\end{equation}

Finally, the weighted sum of cross-products was calculated as:

\begin{equation}
    S_{TR} = \sum_{i=1}^{N} \omega_i \cdot x_i \cdot y_i
\end{equation}

The slope of the regression, corresponding to $-E_\text{a}/R$, was then obtained from:

\begin{equation}
    m = \frac{S \cdot S_{TR} - S_T \cdot S_R}{S \cdot S_{TT} - S_T^2}
\end{equation}

A base uncertainty in $E_\text{a}$ was propagated from the stated measurement precision as:

\begin{equation}
    \sigma_{base}^2 = \frac{R^2 \cdot S}{S \cdot S_{TT} - S_T^2}
\end{equation}

To assess whether the stated measurement uncertainties were consistent with the observed scatter about the fitted line, the reduced chi-squared statistic was calculated as:

\begin{equation}
    \chi^2_{red} = \frac{1}{N-2}\left(S_{RR} - \frac{S_R^2}{S} - \frac{(S \cdot S_{TR} - S_T \cdot S_R)^2}{S(S \cdot S_{TT} - S_T^2)}\right)
\end{equation}

where $N - 2$ represents the degrees of freedom remaining after fitting two parameters (the slope and intercept). Where $\chi^2_{red} > 1$, indicating that the observed scatter exceeded that predicted by the stated $\sigma$, the reported uncertainty was inflated accordingly:

\begin{equation}
    \sigma_{E_\text{a}} = \begin{cases}
        \sigma_{base}, & \text{if } \chi^2_{red} \leq 1 \\
        \sqrt{\chi^2_{red}} \cdot \sigma_{base}, & \text{if } \chi^2_{red} > 1
    \end{cases}
\end{equation}

The resulting activation energy uncertainties were then compared with those originally reported by Huang et al.\ to assess the practical consequences of applying the correct framework.

\section*{Results}

The results are presented in two parts. We first report the findings of the literature audit, establishing the scale of deviation from the correct statistical framework across 57 years of permanganate oxidation kinetics. We then present the reanalysis of the Huang et al.\ dataset, reporting the outcome of the consistency check against the forward model followed by the consequences of applying the correct statistical procedures to both the rate constant and activation energy determinations.

\subsection*{Current Practices in Error Reporting}

We assessed error reporting practices across 330 rate constant measurements compiled by Laszakovits et al.\ \cite{Laszakovits2022} The results of this investigation are shown in Table \ref{tab:error_categories}. 

\begin{table*}[ht]
\centering
\caption{\textbf{Error Reporting Practices in Chemical Kinetics Literature}}
\label{tab:error_categories}
\small
\renewcommand{\arraystretch}{1.6}
\begin{tabular}{p{4cm} p{2cm} p{9cm}}
\hline
\textbf{Category} & \textbf{Count (\%)} & \textbf{Description} \\
\hline
(i) No error reported & 118 (35.8\%) & Neither $k_2$ nor $k_{\text{obs}}$ have a quantified uncertainty \\
(ii) Inaccessible or unverifiable & 73 (22.1\%) & Paper unavailable, or the value reported by Laszakovits et al.\ not found in the paper or its supporting information \\
(iii) Error given, origin not described & 42 (12.7\%) & Uncertainty reported, but no information on its type, the number of repeats, or how it was obtained \\
(iv) Error partially described & 61 (18.5\%) & Some, but not all, of these details--the type of uncertainty, number of repeats, and method--were reported.\\
(v) Error fully described & 36 (10.9\%) & Type of uncertainty, number of repeats, and method all stated \\
(vi) Full uncertainty treatment & 0 (0.0\%) & As (v), with all relevant sources of uncertainty propagated into the reported rate constant \\
\hline
\end{tabular}
\end{table*}

Of the 330 measurements, we find that 118 (35.8\%) are reported with no quantified uncertainty on either $k_2$ or $k_{\text{obs}}$. A further 73 measurements (22.1\%) are inaccessible or unverifiable, meaning the original paper could not be obtained or the reported value could not be located within it. In 42 cases (12.7\%) a numerical uncertainty is reported but no description of how it was derived is provided. In 61 cases (18.5\%) the description is incomplete, giving some but not all of the type of uncertainty, the number of repeats and the method, for example mentioning replicates or averaging without stating whether the uncertainty is a standard deviation or confidence interval. Only 36 measurements (10.9\%) are accompanied by a full description of the uncertainty, and none propagate all relevant sources of uncertainty into the reported rate constant. Furthermore, of the 139 measurements with a reported uncertainty, 17 (12\%) give it only on the pseudo-first-order rate constant $k_{\text{obs}}$ rather than on $k_2$.

\subsection*{Consistency Analysis of Kinetic Parameters from a Representative Study}

\begin{figure}[ht]
    \centering
    \includegraphics[width=1.0\linewidth]{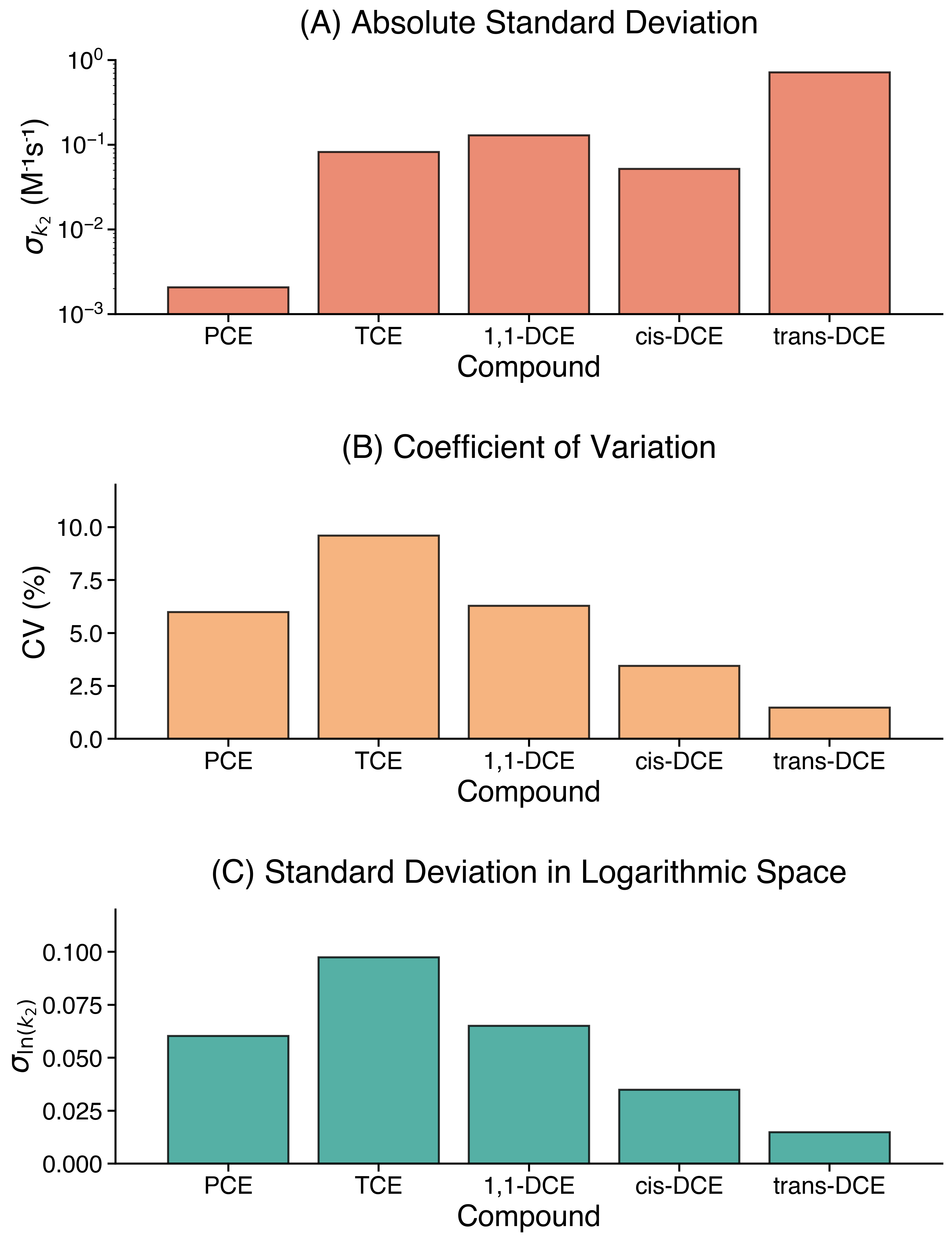}
    \caption{\textbf{Comparison of Variability Metrics Across Five Chlorinated Ethylene Compounds Measured by Huang et al.\ at 20$^\circ$C.} (A) Absolute standard deviation. (B) Coefficient of variation. (C) Standard deviation in log-space. Under a normal distribution, panel A would be expected to remain approximately constant across compounds; under a log-normal distribution, panels B and C would be expected to remain approximately constant.}
    \label{normalVSlognormal}
\end{figure}

We calculated three diagnostic metrics from replicate measurements of $k_2$ at 20$^\circ$C for all five compounds investigated by Huang et al. This can be seen in Figure \ref{normalVSlognormal}.

We find that the absolute standard deviation of $k_2$ varies from $0.002$~M$^{-1}$s$^{-1}$ (PCE) to $0.71$~M$^{-1}$s$^{-1}$ (\textit{trans}-DCE), representing a 355-fold range across compounds (Figure~\ref{normalVSlognormal}(A)). In contrast, CV ranges from 1.5\% to 9.6\%, a 6.4-fold variation (Figure~\ref{normalVSlognormal}(B)), and $\sigma_{\ln,j}$ ranges from 0.015 to 0.097, a 6.5-fold variation (Figure~\ref{normalVSlognormal}(C)). Neither the CV nor $\sigma_{\ln,j}$ is perfectly constant across compounds, with both showing a systematic dependence on reaction rate.

Following this consistency check, we applied the correct estimators derived from the forward model to the Huang et al.\ dataset. For each compound we calculated the geometric mean and asymmetric uncertainty bounds, and report these alongside the arithmetic means and symmetric confidence intervals originally reported by Huang et al.\ in Table \ref{HuangComp}.

\begin{table}[ht]
\centering
\caption{\textbf{Comparison of Rate Constants Reported by Huang et al.\ and Those Obtained Under the Correct Statistical Framework.} Values from Huang et al.\ are reported as an arithmetic mean $\pm$ 95\% confidence interval. Reanalysis values are reported as a geometric mean with asymmetric bounds corresponding to $\pm\sigma_{\ln(k_2)}$ back-transformed to the original space. Significant figures are provided in accordance with those given by the Huang et al.\ study. We report values to three significant figures.}
\label{HuangComp}
\footnotesize
\renewcommand{\arraystretch}{1.6}
\begin{tabular}{p{1.5cm} p{3cm} p{3cm}}
\hline
\textbf{Compound} & \textbf{Huang et al.\ (M$^{-1}$s$^{-1}$)} & \textbf{This work (M$^{-1}$s$^{-1}$)} \\
\hline
PCE              & $0.035 \pm 0.004$ & $0.0345_{-0.00200}^{+0.00210}$ \\
TCE              & $0.80 \pm 0.12$   & $0.850_{-0.0805}^{+0.0889}$   \\
\textit{cis}-DCE & $1.52 \pm 0.05$   & $1.50_{-0.0514}^{+0.0532}$    \\
1,1-DCE          & $2.1 \pm 0.2$     & $2.06_{-0.130}^{+0.138}$      \\
\textit{trans}-DCE & $48.6 \pm 0.9$  & $48.3_{-0.708}^{+0.719}$      \\
\hline
\end{tabular}
\end{table}

\begin{figure}[ht]
    \centering
    \includegraphics[width=1.0\linewidth]{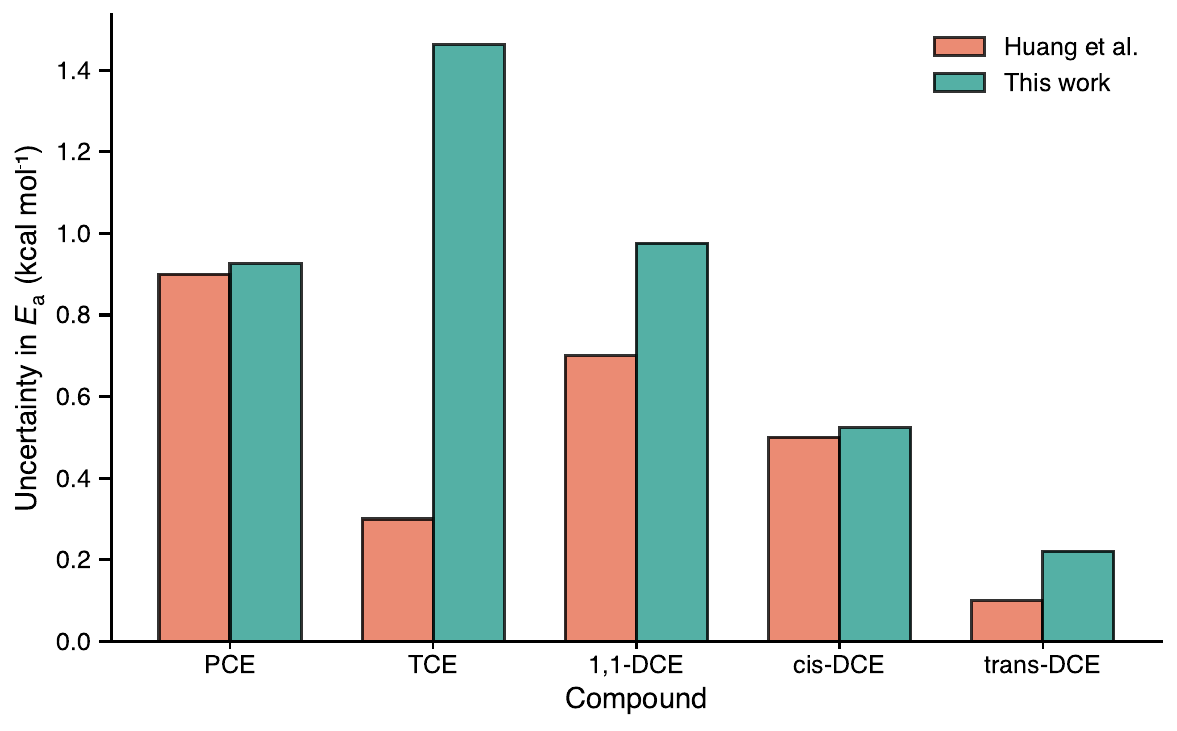}
    \caption{\textbf{Comparison of Activation Energy Uncertainties between the Original Analysis of Huang et al.\ and the Weighted Least Squares Reanalysis Performed Here.} Orange bars show uncertainties derived by propagating stated measurement precision through standard linear regression. Teal bars show uncertainties obtained by weighted least squares regression using empirically derived $\sigma_{\ln}$ values as weights, with a chi-squared correction applied where $\chi^2_{red} > 1$.}
    \label{EaComparison}
\end{figure}

Following this consistency check we determined the activation energy ($E_\text{a}$) for all five compounds using weighted least squares regression. The results are shown in Figure \ref{EaComparison}.

We find that the activation energies obtained here are in close agreement with those reported by Huang et al.\ across all five compounds, with differences of less than 0.5 kcal mol$^{-1}$ in all cases. However, the uncertainties obtained here are larger for all five compounds. The increase is modest for PCE, \textit{cis}-DCE, and 1,1-DCE, but more substantial for TCE, where the uncertainty increases from 0.3 to 1.490 kcal mol$^{-1}$, and for \textit{trans}-DCE, where it increases from 0.1 to 0.221 kcal mol$^{-1}$.

We obtain $\chi^2_{red} > 1$ for only one compound, PCE ($\chi^2_{red} = 1.024$), for which the chi-squared scaling was applied but produced a negligible adjustment. For the remaining four compounds $\chi^2_{red} < 1$ and no scaling was applied.

\section*{Discussion}

In this work, we have presented a forward model describing how kinetic parameters should be distributed. This model is grounded in a theoretical framework that is independent of any particular dataset and provides a basis for determining appropriate statistical procedures for quantifying uncertainty. The framework presented here is not a set of recommendations that researchers may choose to follow or ignore, it is a description of what correct statistical practice looks like given the physical structure of kinetic measurements. 

In this context, the literature audit we performed highlights key problems in how errors are currently reported and quantified. Our findings indicate that not only are 35.8\% of measurements reported with no uncertainty quantification, but that only 10.9\% are accompanied by a full description of how the uncertainty was obtained, and none by a full uncertainty treatment that propagates all relevant sources of uncertainty. We note that this study looked at only a small subset of kinetic data surrounding a very specific reaction, permanganate oxidation kinetics, and thus may not fully reflect the literature surrounding kinetic parameter determination, however, we anticipate that such trends would be seen across the literature, evidenced by the many studies that, like ours, also report a lack of meaningful error quantification \cite{Heberger1987, Goldsmith2013, Wang2015, Vsimon2022}. More fundamentally, the inconsistency in reporting practices across studies suggests that the conceptual framework for thinking correctly about uncertainty in kinetic measurements is largely absent from the literature. We also point out that the papers examined all cover similar measurements for which uncertainty quantification is both possible and tractable, so the absence of errors cannot be attributed to practical limitations.

Given these findings we set out to determine the practical consequences of departing from the correct statistical framework. Our reanalysis of a representative study from Huang et al.\ allowed us to assess this impact. Before applying the correct procedures we first asked whether there was any reason in the data itself to doubt the forward model, that is, whether the data were inconsistent with the multiplicative noise structure the model assumes. Our results show that both CV and $\sigma_{\ln(k_2)}$ are more consistent across compounds than the absolute standard deviation, which is the pattern expected under the forward model. We emphasise that with only five compounds and $n = 5-6$ replicates per compound this check does have limited statistical power and therefore we report this as indicative rather than as any form of confirmation. 

Following this consistency check we apply the correct framework to the Huang et al.\ dataset, and find that the practical consequences are modest. We note that throughout our reanalysis, we use previously derived kinetic parameters, which should therefore have associated uncertainties. Owing to limitations in what data was available, we take these parameters as our starting point. Thus, our analysis is not intended to represent our view of how the data should ideally be reanalysed. For a full example of how we suggest errors should be reported and quantified the reader is referred to the Supporting Information. The geometric means differ only marginally from the arithmetic means originally reported, and the activation energy uncertainties, whilst larger under the correct framework for all five compounds, remain of similar order.  We suggest that this outcome is expected based on the values of $\sigma_{\ln(k_2)}$ being small. When $\sigma_{\ln(k_2)}$ is small the normal and log-normal distributions are approximately equivalent in the region containing most of the probability mass, and inferences about central tendency are robust to the choice of distributional family. The Huang et al.\ dataset falls into this low-$\sigma_{\ln}$ regime. Our analysis further showed that activation energy uncertainties obtained under the correct framework are consistently larger for all five compounds, though only marginally so, with the greatest differences observed for TCE and trans-DCE. This discrepancy arises from a subtle but important distinction between two ways of estimating uncertainty. The approach taken by Huang et al., propagating stated measurement precision through the regression formula, answers the question of how uncertain the activation energy would be if the measurements were exactly as precise as the instruments claimed. The approach taken here, using $\sigma_{\ln(k_2)}$ derived from actual replicate scatter as regression weights, answers the different and more important question of how uncertain the activation energy would actually be given observed rather than assumed measurement variability. That the resulting uncertainties are larger for all five compounds demonstrates that actual variability exceeded stated precision in every case. We also note that while the chi-squared correction was negligible here, it remains a valuable routine diagnostic in datasets where Arrhenius scatter genuinely exceeds measurement variability. Finally, a related inconsistency is also worth noting explicitly. Reporting $k_2$ with symmetric bounds, as was done in the original study by Huang et al., implicitly assumes normality, while performing Arrhenius regression in log-space implicitly assumes log-normality. Both assumptions cannot simultaneously be correct, and the forward model makes clear which is appropriate. This points to a broader inconsistency prevalent across the chemical kinetics literature.

Based on our findings we state that the reason to care about the correct framework does not rest on the findings from this particular dataset but instead rests on the fact that the low-$\sigma_{\ln}$ regime is not universal. As an example, reactions involving highly reactive species present significant measurement challenges because reactants and products participate in competing secondary reactions that complicate the interpretation of experimental data. Reported rate constants for such reactions can span orders of magnitude across studies \cite{Speak2023}. This is precisely the domain where the equivalence between normal and log-normal breaks down and where symmetric bounds will routinely extend below zero. Interestingly, individual studies measuring such reactions typically report consistency with normality, this may be due to the fact that within a single study, the observed spread is small, and in that low-$\sigma_{\ln}$ regime the distributional assumption matters little. It is only when measurements are pooled across many studies that the true spread becomes apparent, spanning orders of magnitude and suggesting the underlying distribution may not be normal. No individual study ever samples enough of that full spread to detect the departure from normality and indeed, if one did, even a modest number of measurements would suffice to flag it, since larger $\sigma_{\ln}$ makes distributional differences easier to detect. This illustrates why establishing the value of $\sigma_{\ln}$ at which distributional assumptions begin to matter, and how many samples would be needed to detect this, is a practically important question and one that cannot be answered from within any single study alone. These considerations are particularly acute for chemical processes in atmospheric chemistry, astrochemistry, and prebiotic chemistry where reactions are kinetically complex. 

The findings presented here provide both empirical evidence of the scale of the problem and demonstrate that even carefully conducted studies following standard practice can produce overconfident uncertainty estimates and internally inconsistent statistical treatments. As the scale and automation of kinetic data aggregation grows, these errors will propagate silently through predictive frameworks in ways that are increasingly difficult to detect or correct retrospectively. This is not merely a concern for aqueous-phase kinetic parameters, where no evaluated database exists and normality is commonly assumed, even in the gas-phase community, where expert evaluation panels have implicitly adopted the correct distributional framework \cite{Atkinson2006, Burkholder2020}, the consequences of informal uncertainty assignment are significant. Newsome and Evans \cite{Newsome2017} demonstrated that propagating JPL and IUPAC uncertainty factors through a global atmospheric chemistry model yields uncertainties on policy-relevant atmospheric composition metrics that are comparable to or exceed the inter-model spread typically used as the benchmark for uncertainty in climate assessments. This arises because all models draw from the same evaluated rate constants, meaning inter-model comparisons are structurally blind to this source of uncertainty. They also state that the estimate is itself a lower bound, derived from uncertainty factors that are assigned through expert judgement rather than formal statistical analysis, and that are built on primary measurements whose uncertainty characterisation is, as demonstrated here, inconsistent and often absent. The root cause is not a lack of awareness of uncertainty, but a lack of any agreed framework for characterising it correctly at the point of measurement -- before it enters databases, before it is evaluated by expert panels, and before it propagates into models whose outputs inform policy. Addressing this requires not just better reporting habits but adoption of a correct statistical framework from the outset. The recommendations that follow are intended to make this tractable in practice.

\section*{Recommendations}

The following recommendations follow directly from the forward model derived in this work and are intended to make correct statistical practice accessible for experimentalists. They address the identification of uncertainty sources, the specification of a forward model, the reporting of uncertainties in a form consistent with that model, the correct treatment of regression-derived parameters, and the minimum experimental design required to support these procedures.

Our first recommendation is to \textbf{identify and propagate all instrument uncertainties explicitly (R1)}. The forward model requires an estimate of the uncertainty, and this estimate is grounded in the uncertainties of the instruments that contribute to the final result.  Before taking measurements, the stated tolerances of all instruments should therefore be accounted for. In solution-phase kinetics, for example, errors in mass measurements affect the calculated number of moles and therefore the prepared concentrations. Errors in volume measurements influence dilution factors and likewise affect concentrations. In addition, the precision of the analytical instrument used to monitor the reaction directly impacts the measured reaction rate. We recommend computing two quantities from these sources. The first is a statistical $\sigma$, obtained by combining all relative contributions in quadrature (taking the square root of the sum of squares of the individual uncertainties) under the assumption that the errors are independent and randomly distributed. The second is a worst-case bound $\delta$, obtained by summing the tolerances directly, assuming all errors are simultaneously maximally unfavourable. These two quantities should be maintained as separate tracks throughout the analysis.

Second, we recommend that authors \textbf{specify a forward model based on the physical structure of their measurement process (R2)}. The forward model derived in this work provides a template that will be appropriate for many solution-phase kinetic measurements where uncertainty enters multiplicatively. For other measurement processes the same approach should be followed. The forward model should describe how errors enter the measurement based on the physical structure of the instruments and calculations involved, not on the data themselves. Once specified, the data can be checked for consistency with the model using the diagnostic metrics described in this work. The absolute standard deviation, CV, and log-transformed standard deviation should be compared across replicate measurements, and Q-Q plots constructed for both untransformed and log-transformed data where sample sizes permit. We emphasise that consistency checks should be performed not to validate the forward model, which is justified on theoretical grounds, but to identify anomalous datasets and to assess whether uncertainty is sufficiently constant across conditions to justify uniform regression weights.

Third, we recommend \textbf{reporting uncertainties in a way that is consistent with the forward model (R3)}. As established in this work, the correct form of reported uncertainty depends on which parameter is being reported and follows directly from the forward model. Rate constants, rates, and pre-exponential factors are log-normally distributed and should be reported as geometric means with asymmetric bounds. Reaction orders and activation energies are normally distributed and should be reported with symmetric bounds.

Fourth, we recommend \textbf{performing weighted least squares rather than ordinary least squares regression, and computing the reduced chi-squared statistic for all regression-based analyses (R4)}. As the forward model states, the maximum likelihood estimator for any regression in which the dependent variable is a log-transformed kinetic quantity is weighted least squares with weights $\omega = 1/\sigma^2_{\ln}$. In addition, the reduced chi-squared statistic should be calculated and reported alongside all regression results, and where it exceeds unity the uncertainty in the fitted parameter should be scaled by $\sqrt{\chi^2_{red}}$ as a conservative correction for unmodelled variability.

Following each of these recommendations, \textbf{both a statistical estimate and an absolute range should be reported (R5)}. For every reported quantity two parallel outputs should be maintained. The statistical estimate is the mean with its associated uncertainty bounds as determined by the forward model. The absolute range is obtained by exhaustive combination of all instrument tolerance bounds and replicate extremes, making no distributional assumptions and providing a conservative bound that is always valid regardless of whether the forward model holds. The statistical uncertainty is more informative but depends on assumptions that should be stated explicitly.

Finally, we recommend \textbf{a minimum of three independent replicate measurements (R6)}. The forward model requires estimation of uncertainty from replicate measurements, and a single measurement provides no basis for this estimation, nor for computing a range or performing any consistency checks. Replicates should be independent preparations rather than technical repeats of the same prepared solution, since only independent replicates sample the full uncertainty of the measurement process.

An example of how these recommendations can be applied in practice can be found in the Supporting Information.

\section*{Conclusions}

Chemical kinetics has long supported predictive modelling, however as kinetic data are increasingly aggregated and reused, the quality and consistency of uncertainty reporting has become more consequential. This work has shown that the foundation for correct uncertainty quantification already exists in the physical structure of kinetic measurements themselves, and that deriving it explicitly reveals both what correct practice looks like and how far the literature currently falls short. Our example literature audit and reanalysis of a representative chemical kinetics study indicate that current methods of error reporting and quantification are often absent or, when reported, incorrect and/or inconsistent. In light of our findings, we provide clear recommendations for reporting and quantifying errors. Following these guidelines will improve the reliability of future models and tools and help ensure that scientific conclusions remain robust, even as automated analyses and large language models become increasingly involved in the aggregation of chemical kinetics data and the outcomes of predictive models.

\section*{Author contributions}
\textbf{S. B. White}: conceptualisation, data curation, formal analysis, investigation, methodology, visualisation, writing -- original draft, 
writing -- review \& editing. 
\textbf{P. B. Rimmer}: conceptualisation, writing -- review. 
\textbf{M. Cramner}: writing -- review. 
\textbf{A. T. Archibald}: writing -- review.

\section*{Conflicts of interest}

There are no conflicts to declare.

\section*{Data availability}

The data supporting this article, including all spreadsheets and associated files, are available on GitHub at \href{https://github.com/astroskyla/Quantifying-Errors-in-Chemical-Kinetics.}\href{https://github.com/astroskyla/Quantifying-Errors-in-Chemical-Kinetics.} An SI is also provided which includes a derivation of the coefficient of variation and it's relations as well as a worked example of how to report and quantify errors following the recommendations outlined in this work. For a full description of the literature audit the reader is referred to the GitHub.

\section*{Acknowledgements}
SBW acknowledges support from the Leverhulme Centre for Life in the Universe (Grant G112026, Seed Fund, Project KKZA/253), the Department of Physics (University of Cambridge), and Newnham College (University of Cambridge). ATA acknowledges funding from NERC NE/W009412/1. SBW thanks Kaisey Mandel for his feedback on the statistical framework and Jay Green for many useful discussions. 

%%%END OF MAIN TEXT%%%



%%%REFERENCES%%%
\bibliography{rsc} %You need to replace "rsc" on this line with the name of your .bib file

@article{Hossain2012,
  title={Oxidation of chlorophenols in aqueous solution by excess potassium permanganate},
  author={Hossain, SM Ghausul and McLaughlan, Robert G},
  journal={Water, Air, \& Soil Pollution},
  volume={223},
  number={3},
  pages={1429--1435},
  year={2012},
  publisher={Springer}
}

@article{Grinberg2026,
  title={Predictive Chemical Kinetic Modeling: Where We Succeed, Where We Struggle, and What Comes Next},
  author={Grinberg Dana, Alon and Van Geem, Kevin M and Cavallotti, Carlo and Green, William H},
  journal={ACS Engineering Au},
  year={2026},
  publisher={ACS Publications}
}

@article{Abel2017,
  title={Using chemical reaction kinetics to predict optimal antibiotic treatment strategies},
  author={Abel zur Wiesch, Pia and Clarelli, Fabrizio and Cohen, Ted},
  journal={PLoS computational biology},
  volume={13},
  number={1},
  pages={e1005321},
  year={2017},
  publisher={Public Library of Science San Francisco, CA USA}
}

@article{Wang2007,
  title={A model of biogeochemical cycles of carbon, nitrogen, and phosphorus including symbiotic nitrogen fixation and phosphatase production},
  author={Wang, Y-P and Houlton, BZ and Field, CB},
  journal={Global Biogeochemical Cycles},
  volume={21},
  number={1},
  pages={},
  year={2007},
  publisher={Wiley Online Library}
}

@article{Chen2005,
  title={Kinetics of the oxidation of MCRR by potassium permanganate},
  author={Chen, Xiaoguo and Xiao, Bangding and Liu, Jiantong and Fang, Tao and Xu, Xiaoqing},
  journal={Toxicon},
  volume={45},
  number={7},
  pages={911--917},
  year={2005},
  publisher={Elsevier}
}

@article{Hesterberg2015,
  title={What teachers should know about the bootstrap: Resampling in the undergraduate statistics curriculum},
  author={Hesterberg, Tim C},
  journal={The american statistician},
  volume={69},
  number={4},
  pages={371--386},
  year={2015},
  publisher={Taylor \& Francis}
}

@article{Laszakovits2022,
  title={Permanganate oxidation of organic contaminants and model compounds},
  author={Laszakovits, Juliana R and Kerr, Adaline and MacKay, Allison A},
  journal={Environmental Science \& Technology},
  volume={56},
  number={8},
  pages={4728--4748},
  year={2022},
  publisher={ACS Publications}
}

@article{Bazyleva2021,
  title={Good reporting practice for thermophysical and thermochemical property measurements (IUPAC Technical Report)},
  author={Bazyleva, Ala and Abildskov, Jens and Anderko, Andrzej and Baudouin, Olivier and Chernyak, Yury and de Hemptinne, Jean-Charles and Diky, Vladimir and Dohrn, Ralf and Elliott, J Richard and Jacquemin, Johan and others},
  journal={Pure and applied chemistry},
  volume={93},
  number={2},
  pages={253--272},
  year={2021},
  publisher={de Gruyter}
}

@article{Bruno2021,
  title={FAIR and open data in science: the opportunity for IUPAC},
  author={Bruno, Ian and Coles, Simon and Koch, Wolfram and McEwen, Leah and Meyers, Fabienne and Stall, Shelley},
  journal={Chemistry International},
  volume={43},
  number={3},
  pages={12--16},
  year={2021},
  publisher={De Gruyter}
}

@article{Shaw-2023,
  title={Chemical data evaluation: general considerations and approaches for IUPAC projects and the chemistry community (IUPAC Technical Report)},
  author={Shaw, David G and Bruno, Ian and Chalk, Stuart and Hefter, Glenn and Hibbert, David Brynn and Hutchinson, Robin A and Magalh{\~a}es, M Clara F and Magee, Joseph and McEwen, Leah R and Rumble, John and others},
  journal={Pure and applied chemistry},
  volume={95},
  number={10},
  pages={1107--1120},
  year={2023},
  publisher={De Gruyter}
}

@manual{eurachem1995_quantifying_uncertainty,
  title        = {Quantifying Uncertainty in Analytical Measurement},
  author       = {{EURACHEM Working Group on Uncertainty in Chemical Measurement}},
  year         = {1995},
  note         = {Reprinted 2022. Available at: \url{https://eurachem.org/images/stories/Guides/pdf/QUAM95_2022.pdf}},
  organisation = {EURACHEM},
  address      = {Teddington, United Kingdom}
}

@incollection{Verma2025,
  title={Relevance of Good Laboratory Practices in Conformity Assessment},
  author={Verma, Mitra Sen and Gupta, Rajat},
  booktitle={Handbook of Quality System, Accreditation and Conformity Assessment},
  pages={1--19},
  year={2025},
  publisher={Springer}
}

@misc{Hartland2020,
  title={Statistical analysis of physical chemistry data: errors are not mistakes},
  author={Hartland, Gregory V},
  journal={The Journal of Physical Chemistry C},
  volume={124},
  number={11},
  pages={5889--5892},
  year={2020},
  publisher={ACS Publications}
}

@article{Vsimon2022,
  title={Frequent flaws encountered in the manuscripts of kinetic papers: P. {\v{S}}imon et al.},
  author={{\v{S}}imon, Peter and Dubaj, Tibor and Cibulkov{\'a}, Zuzana},
  journal={Journal of Thermal Analysis and Calorimetry},
  volume={147},
  number={18},
  pages={10083--10088},
  year={2022},
  publisher={Springer}
}

@techreport{JCGM100:2008,
  author      = {{BIPM} and {IEC} and {IFCC} and {ILAC} and {ISO} and {IUPAC} and {IUPAP} and {OIML}},
  title       = {Evaluation of Measurement Data --- {Guide} to the Expression of Uncertainty in Measurement},
  institution = {Joint Committee for Guides in Metrology},
  year        = {2008},
  number      = {JCGM 100:2008},
  doi         = {10.59161/JCGM100-2008E}
}

@book{Taylor2022,
  author    = {Taylor, John R.},
  title     = {An Introduction to Error Analysis: The Study of Uncertainties in Physical Measurements},
  edition   = {3},
  publisher = {The MIT Press},
  year      = {2022},
  isbn      = {9781940380087}
}

@article{Fernandez2006,
  title={Modeling the kinetics of bimolecular reactions},
  author={Fern{\'a}ndez-Ramos, Antonio and Miller, James A and Klippenstein, Stephen J and Truhlar, Donald G},
  journal={Chemical reviews},
  volume={106},
  number={11},
  pages={4518--4584},
  year={2006},
  publisher={ACS Publications}
}

@article{Huang2001,
  title={Oxidation of chlorinated ethenes by potassium permanganate: a kinetics study},
  author={Huang, Kun-Chang and Hoag, George E and Chheda, Pradeep and Woody, Bernard A and Dobbs, Gregory M},
  journal={Journal of Hazardous Materials},
  volume={87},
  number={1-3},
  pages={155--169},
  year={2001},
  publisher={Elsevier}
}

@article{Cvetanovic1979,
  title={Evaluations of the mean values and standard errors of rate constants and their temperature coefficients},
  author={Cvetanovic, RJ and Singleton, DL and Paraskevopoulos, G},
  journal={Journal of Physical Chemistry},
  volume={83},
  number={1},
  pages={50--60},
  year={1979},
  publisher={ACS Publications}
}

@article{Pardue1989,
  title={Kinetic aspects of analytical chemistry},
  author={Pardue, Harry L},
  journal={Analytica Chimica Acta},
  volume={216},
  pages={69--107},
  year={1989},
  publisher={Elsevier}
}

@article{Wang2015,
  title={Combustion kinetic model uncertainty quantification, propagation and minimization},
  author={Wang, Hai and Sheen, David A},
  journal={Progress in Energy and Combustion Science},
  volume={47},
  pages={1--31},
  year={2015},
  publisher={Elsevier}
}

@article{Goldsmith2013,
  title={Uncertainty propagation in the derivation of phenomenological rate coefficients from theory: A case study of n-propyl radical oxidation},
  author={Goldsmith, C Franklin and Tomlin, Alison S and Klippenstein, Stephen J},
  journal={Proceedings of the Combustion Institute},
  volume={34},
  number={1},
  pages={177--185},
  year={2013},
  publisher={Elsevier}
}

@article{Komp2022,
  title={Progress towards machine learning reaction rate constants},
  author={Komp, Evan and Janulaitis, Nida and Valleau, St{\'e}phanie},
  journal={Physical Chemistry Chemical Physics},
  volume={24},
  number={5},
  pages={2692--2705},
  year={2022},
  publisher={Royal Society of Chemistry}
}

@article{Johnson2022,
  title={RMG database for chemical property prediction},
  author={Johnson, Matthew S and Dong, Xiaorui and Grinberg Dana, Alon and Chung, Yunsie and Farina Jr, David and Gillis, Ryan J and Liu, Mengjie and Yee, Nathan W and Blondal, Katrin and Mazeau, Emily and others},
  journal={Journal of Chemical Information and Modeling},
  volume={62},
  number={20},
  pages={4906--4915},
  year={2022},
  publisher={ACS Publications}
}

@article{Heberger1987,
  title={On the errors of Arrhenius parameters and estimated rate constant values},
  author={H{\'e}berger, K{\'a}roly and Kem{\'e}ny, S{\'a}ndor and Vid{\'o}czy, Tam{\'a}s},
  journal={International journal of chemical kinetics},
  volume={19},
  number={3},
  pages={171--181},
  year={1987},
  publisher={Wiley Online Library}
}

@article{Yano2022,
  title={The case for data science in experimental chemistry: examples and recommendations},
  author={Yano, Junko and Gaffney, Kelly J and Gregoire, John and Hung, Linda and Ourmazd, Abbas and Schrier, Joshua and Sethian, James A and Toma, Francesca M},
  journal={Nature Reviews Chemistry},
  volume={6},
  number={5},
  pages={357--370},
  year={2022},
  publisher={Nature Publishing Group UK London}
}

@inproceedings{Franco2015,
  title={Uncertainty propagation in biomedical models},
  author={Franco, Andrea and Correia, Marco and Cruz, Jorge},
  booktitle={Conference on Artificial Intelligence in Medicine in Europe},
  pages={166--171},
  year={2015},
  organization={Springer}
}

@article{Andersson2021,
  title={Mechanisms for log normal concentration distributions in the environment},
  author={Andersson, August},
  journal={Scientific reports},
  volume={11},
  number={1},
  pages={16418},
  year={2021},
  publisher={Nature Publishing Group UK London}
}

@article{Ciriminna2024,
  title={Reproducibility in chemistry research},
  author={Ciriminna, Rosaria and Angellotti, Giuseppe and Petri, Giovanna Li and Pagliaro, Mario},
  journal={Heliyon},
  volume={10},
  number={14},
  pages={},
  year={2024},
  publisher={Elsevier}
}

@article{Loison2015,
  title={The neutral photochemistry of nitriles, amines and imines in the atmosphere of Titan},
  author={Loison, JC and H{\'e}brard, Eric and Dobrijevic, M and Hickson, KM and Caralp, F and Hue, V and Gronoff, G and Venot, Olivia and B{\'e}nilan, Y},
  journal={Icarus},
  volume={247},
  pages={218--247},
  year={2015},
  publisher={Elsevier}
}

@article{Hebrard2015,
  title={Determining predictive uncertainties and global sensitivities for large parameter systems: A case study for n-butane oxidation},
  author={H{\'e}brard, Eric and Tomlin, Alison S and Bounaceur, Roda and Battin-Leclerc, Fr{\'e}d{\'e}rique},
  journal={Proceedings of the Combustion Institute},
  volume={35},
  number={1},
  pages={607--616},
  year={2015},
  publisher={Elsevier}
}

@article{Motulsky2025,
  title={Analyzing lognormal data: A nonmathematical practical guide},
  author={Motulsky, Harvey J and Head, Trajen and Clarke, Paul BS},
  journal={Pharmacological Reviews},
  volume={77},
  number={3},
  pages={100049},
  year={2025},
  publisher={Elsevier}
}

@article{Wakelam2010,
  title={Sensitivity analyses of dense cloud chemical models},
  author={Wakelam, Valentine and Herbst, E and Le Bourlot, J and Hersant, F and Selsis, Franck and Guilloteau, S},
  journal={Astronomy \& Astrophysics},
  volume={517},
  pages={A21},
  year={2010},
  publisher={EDP Sciences}
}

@article{Speak2023,
  title={New Measurements and calculations on the kinetics of an old reaction: OH+ HO2-> H2O+ O2},
  author={Speak, Thomas H and Blitz, Mark A and Medeiros, Diogo J and Seakins, Paul W},
  journal={JACS Au},
  volume={3},
  number={6},
  pages={1684--1694},
  year={2023},
  publisher={ACS Publications}
}

@article{White2024b,
  title={Do-Nothing Prebiotic Chemistry: Chemical Kinetics as a Window into Prebiotic Plausibility},
  author={White, Skyla B and Rimmer, Paul B},
  journal={Accounts of Chemical Research},
  volume={58},
  number={1},
  pages={1--10},
  year={2024},
  publisher={ACS Publications}
}

@article{Carvalho2019,
  title={Temperature dependence of rate processes beyond Arrhenius and Eyring: Activation and Transitivity},
  author={Carvalho-Silva, Valter H and Coutinho, Nayara D and Aquilanti, Vincenzo},
  journal={Frontiers in chemistry},
  volume={7},
  pages={380},
  year={2019},
  publisher={Frontiers Media SA}
}

@article{Atkinson2006,
  title={Evaluated kinetic and photochemical data for atmospheric chemistry: Volume II--gas phase reactions of organic species},
  author={Atkinson, R and Baulch, DL and Cox, RA and Crowley, JN and Hampson, RF and Hynes, RG and Jenkin, ME and Rossi, MJ and Troe, Juergen and Subcommittee, IUPAC},
  journal={Atmospheric chemistry and physics},
  volume={6},
  number={11},
  pages={3625--4055},
  year={2006},
  publisher={Copernicus Publications G{\"o}ttingen, Germany}
}

@article{Newsome2017,
  title={Impact of uncertainties in inorganic chemical rate constants on tropospheric composition and ozone radiative forcing},
  author={Newsome, Ben and Evans, Mat},
  journal={Atmospheric Chemistry and Physics},
  volume={17},
  number={23},
  pages={14333--14352},
  year={2017},
  publisher={Copernicus Publications G{\"o}ttingen, Germany}
}

@techreport{Burkholder2020,
  title={Chemical kinetics and photochemical data for use in atmospheric studies; evaluation number 19},
  author={Burkholder, JB and Sander, SP and Abbatt, JPD and Barker, JR and Cappa, C and Crounse, JD and Dibble, TS and Huie, RE and Kolb, CE and Kurylo, MJ and others},
  year={2020},
  institution={Pasadena, CA: Jet Propulsion Laboratory, National Aeronautics and Space~…}
}

@techreport{Possolo2015,
  author      = {Possolo, Antonio},
  title       = {Simple Guide for Evaluating and Expressing the Uncertainty of {NIST} Measurement Results},
  institution = {National Institute of Standards and Technology},
  year        = {2015},
  number      = {NIST TN 1900},
  doi         = {10.6028/NIST.TN.1900},
  url         = {https://nvlpubs.nist.gov/nistpubs/TechnicalNotes/NIST.TN.1900.pdf}
}

@article{Mellouki2021,
  title={Evaluated kinetic and photochemical data for atmospheric chemistry: volume VIII--gas-phase reactions of organic species with four, or more, carbon atoms (≥ C 4)},
  author={Mellouki, Abdelwahid and Ammann, Markus and Cox, R Anthony and Crowley, John N and Herrmann, Hartmut and Jenkin, Michael E and McNeill, V Faye and Troe, J{\"u}rgen and Wallington, Timothy J},
  journal={Atmospheric Chemistry and Physics},
  volume={21},
  number={6},
  pages={4797--4808},
  year={2021},
  publisher={Copernicus GmbH}
}

@article{Mcgillen2020,
  title={Database for the kinetics of the gas-phase atmospheric reactions of organic compounds},
  author={McGillen, Max R and Carter, William PL and Mellouki, Abdelwahid and Orlando, John J and Picquet-Varrault, B{\'e}n{\'e}dicte and Wallington, Timothy J},
  journal={Earth System Science Data},
  volume={12},
  number={2},
  pages={1203--1216},
  year={2020},
  publisher={Copernicus Publications G{\"o}ttingen, Germany}
}
\bibliographystyle{rsc} %the RSC's .bst file

\end{document}